# Dynamic Cavitation Inception by Wave Propagation Across Solid–Fluid Interface with Varying Solid Surface Wettability

**Tomohisa Kojima[1]**
Department of Precision Mechanics, Chuo University
1-13-27, Kasuga, Bunkyo-ku, Tokyo, 112-8551, Japan
kojima.31k@g.chuo-u.ac.jp
ASME Member

**Kazuaki Inaba**
Department of Transdisciplinary Science and Engineering, Tokyo Institute of Technology
2-12-1-I6-5, Ookayama, Meguro-ku, Tokyo, 152-8550, Japan
inaba.k.ag@m.titech.ac.jp
ASME Member

**ABSTRACT**

*Fluid-structure interaction (FSI) problems are important because they may induce serious damage to structures. In some FSI problems, the interaction mechanism is strongly dependent on the wave propagation across the solid–fluid interface. In this study, we attempted a quantitative evaluation of the effect of the solid surface wettability on the wave propagation across the solid–fluid interface with FSI in the case of longitudinal wave propagation vertically towards the interface. During the experiments, while the water was continuously compressed by the solid buffer motion, cavitation bubbles appeared being originated from the buffer–water interface as a result of the transmitted tensile wave propagating across the interface in a cycle. It was confirmed that interfacial boundary condition as wettability could change the wave transmission behavior owing to changes in the cavitation occurrence. It was also confirmed that the worse the wettability, the severer the cavitation intensity, and the greater the difference between the energy lost by the buffer and the energy stored in the water. Consequently, the effect of the cavitation inception on the wave propagation*

---

[1]Tomohisa Kojima is a research associate of the Department of Precision Mechanics, Faculty of Science and Engineering, Chuo University, Japan. He received his doctoral degree in engineering from Tokyo Institute of Technology, Japan, in 2017. He started his career as an assistant professor at Meiji University, Japan, in the same year. He has been working in the current position since 2019.

*at the solid–fluid interface with FSI could be quantitatively evaluated by considering the energy transferred from the solid to the water.*

# 1 INTRODUCTION

Fluid-structure interaction (FSI) problems with wave propagation are important because they may induce serious damage to structures. These problems include the response of water hammers in pipelines [1–5], pulsating blood vessels and hearts [6–8], and the response of structures subjected to underwater explosions [9,10]. In these FSI problems, the interaction mechanism is strongly dependent on the wave propagation across the solid–fluid interface [11,12]. Therefore, it is crucial to understand the wave propagation mechanics across the solid–fluid interface to ensure the safety and reliability of fluid machines. However, the mechanics of wave propagation across the solid–fluid interface with structural motion have not yet been clarified. In the case of a water hammer, a low-pressure surge can be generated in addition to high pressure. The generated low pressure vaporizes the water, thereby inducing small bubbles that are distributed in the pipe, which is known as cavitation.

Numerous studies have been conducted regarding cavitation with FSI. An FSI experiment with transient column separation in a closed straight pipe was conducted by Fan and Tijsseling [13]. It was observed that stress waves in the pipe wall induced distributed vaporous cavitation. Tijsseling and Vardy experimentally and numerically investigated the effects of cavitation on a water hammer with an unrestrained single-elbow pipe and unrestrained single branch [14,15]. Inaba and Shepherd clarified when and where large cavitation formations occurred with pressure wave propagation in the axial direction using a closed straight pipe [16]. Schiffer et al. analytically modeled the response of a

solid–fluid interface on blast loading with cavitation occurrence [17,18]. More recently, Veilleux et al. investigated the pressure wave characteristic with the cavitation inception and collapse in their studies of wave propagation with an air gap between the solid and fluid [19,20]. Owing to these studies, FSI with the cavitation effect has been clarified and modeled. In recent years, accurate numerical models have been proposed for fluid-structures interaction due to water hammer, including column separation and cavitation [21-23]. However, the effect of the cavitation initiation and growth on the FSI at the solid–fluid interface has not yet been clarified. It is possible that the cavitation will not expand enough to affect the FSI phenomenon, as is the case with an initial crack not growing in the fracture mechanics of solid materials [24].

In general, the condition of the solid–fluid interface is not considered in FSI problems with wave propagation. The impact of liquid mercury droplets on a solid surface was studied [25]. It was reported that the surface tension of the liquid droplets exerted greater impulsive pressure at the solid surface compared to the case when it was absent. Therefore, the surface conditions of the solid medium at the solid-liquid interface, such as the surface roughness and wettability, may affect the impulsive pressure owing to the FSI, as these surface conditions of the solid are related to the liquid surface tension at the solid-liquid interface [26]. Moreover, it was observed that instantaneous cavitation originated at the solid–fluid interface as a result of tensile stress wave propagation from a solid medium, which could affect the wave propagation across the interface [27,28]. When cavitation bubbles appear, liquid–gas, and solid-gas interfaces are generated and grow [29]. Because the generation and growth of these interfaces depend on the surface energy of the solid and fluid [30,31], it can be considered that the solid surface condition at the solid–fluid interface affects the wave propagation at the interface. Although an experiment was previously carried out to investigate the effect on the wave propagating across the solid–fluid interface by changing the solid surface state, the quantitative evaluation was not sufficient [32].

The wave reflection and transmission at dissimilar interfaces can be explained by the theory of acoustic impedance. However, there has been no theory for solid–fluid or fluid-solid interfaces in acoustic theory. In the previous study, a one-dimensional theory that explains wave propagation across the solid–fluid interface has been investigated based on acoustic theory [27,12]. Besides, the effect of the surface conditions at the interface on wave transmission was also investigated [32]. However, the quantitative evaluation of the effect of the surface condition of the solid and cavitation have been insufficient.

Therefore, in this study, we attempted to evaluate the effect of surface wettability of the solid medium on the wave propagation across the solid–fluid interface with FSI quantitatively in the case of longitudinal wave propagation vertically towards the interface. The results of this work may provide suggestions for wave conditioning or attenuation in water hammer events.

## 2 METHODS

### 2.1 Experimental setup

Figure 1 shows a schematic of the experimental setup used in this study. A buffer made of a 300 mm long aluminum (Al) rod was placed on the upper end of a 1000 mm long polycarbonate standing pipe filled with water and closed the pipe. The diameter of the buffer, the inner diameter of the pipe, and the thickness of the pipe are 50 mm, 52 mm, and 4 mm, respectively. When the buffer is subjected to impact loading by free-falling steel projectile, stress waves are generated in it; these waves propagated through the buffer and across the interface of the buffer and water in the pipe. Thus, the stress wave in the solid propagated to the water in the pipe as a pressure wave. During the experiments, the axial strains of the buffer were measured with the strain gauges at g1 and g2; the hoop strains of the water-filled pipe were measured at g3–g11 as shown in Fig. 1. The motions of the buffer and water close to their interface were observed using a high-speed video camera (VW-9000, KEYENCE) at a frame rate of 230000 fps and shutter speed of 1/900000. The drop height of the projectile above the top of the buffer was varied as 100, 200, and 300 mm, and the experiment was repeated five times for each condition.

**2.2 Experimental conditions**

As the experimental condition, the Al buffers with different surface wettabilities on their edge were used. To change the surface wettability, plasma-treated Al foil tape (425, 3MTM treated with YHS-R, SAKIGAKE-Semiconductor Co., Ltd.) was attached to the bottom of the buffer. The Al foil tape used in the experiment was 0.11 mm-thick and it was plasma-treated for 180 s; the plasma treatment time was determined from the technical data of SAKIGAKE-Semiconductor Co., Ltd. [33]. A buffer with untreated Al foil tape was also prepared as a reference. The surface wettability is quantitatively assessed by contact angle with a liquid droplet. Figure 2 and Table 1 show the contact angles with water droplets measured for each surface condition in the present study. These measurements utilized the half-angle method [34]. We also measured the surface roughness in each condition using a laser microscope (VK-9710, KEYENCE). The wettability improved in the following order: buffer edge, attached untreated Al tape, and attached plasma-treated

Al tape. Correspondingly, the contact angles decreased in the same order, while the surface roughness was similar for each condition.

During the experiment, the dissolved oxygen amount was measured using a dissolved oxygen meter (Portavo 907 Multi with the sensor head SE340, HORIBA Advanced Techno, Co., Ltd.), and was found to be 8.24 mg/L at the water temperature of 24.6 °C. This value is almost equivalent to the saturated dissolved oxygen in normal water.

**2.3 Evaluation method**

The effect of the cavitation intensity on the wave propagation with the difference in wettability of the solid surface was evaluated by the energy transmission quantity.

*2.3.1. Theoretical model of kinetic energy loss of solid buffer*

Assuming that the kinetic energy of the buffer is $K_s(t)$ at time t, the kinetic energy $K_{s,lost}$ lost by the buffer because of the energy transmission to water is expressed as follows:

$$K_{s,lost} = K_s(t) - K_s(0) \tag{1}$$

A factor for the buffer to lose its kinetic energy is friction between the inner tube wall and the O-rings. In this study, the contact surfaces between them were lubricated with grease, so the friction should be small. Assuming that the buffer is rigid,

$$K_s(t) = \frac{1}{2} m_b \{v(t)\}^2 \tag{2}$$

where $m_b$ is the buffer mass and $v(t)$ is the buffer velocity. As the buffer is subjected to resistance $p(t)$ from water, according to Newton's equation of motion:

$$m_b \frac{d[v(t)]}{dt} = p(t) A_b \tag{3}$$

According to Joukowsky's equation [35],

$$p(t) = \rho_w c_K v(t) \tag{4}$$

where $\rho_w$ is the water density and $c_K$ is the propagation speed of the pressure wave. The propagation speed of the pressure wave propagating in the axial direction in the water in the pipe becomes slower than the usual speed of sound owing to the FSI with the pipe wall. The wave speed of a pressure wave propagating in the axial direction in an elastic circular pipe is expressed by the following Korteweg velocity [36]:

$$c_K = \frac{c_w}{\sqrt{1 + 2RK/aE_{tube}}} \tag{5}$$

where $c_w$ is the sound speed of the water, $a$ is the pipe wall thickness, $R$ is the representative radius, $K$ is the water volume modulus, and $E_{pipe}$ is Young's modulus of the pipe material. Using Eqs. (4) and (3), $v(t)$ can be written as follows:

$$v(t) = v(0)\exp\left(-\frac{\rho_w c_K A_b}{m_b} t\right) \tag{6}$$

Therefore, from Eq. (1),

$$K_{s,lost} = \frac{1}{2} m_b \left[v(0)\right]^2 \left[1 - \exp\left(-\frac{\rho_w c_K A_b}{m_b} t\right)\right] \tag{7}$$

The initial velocity $v(0)$ of the buffer was derived from momentum conservation. If the projectile is assumed to be a rigid body as well as the buffer, from momentum conservation,

$$m_p V_0 = m_p V_1 + m_b v(0) \tag{8}$$

where $m_p$ is the projectile mass, $V_0$ and $V_1$ are the projectile velocity before and after the collision, respectively. The relationship between $V_0$, $V_1$ and $v(0)$ can be expressed by Newton's law of restitution. According to Newton's law of restitution, the coefficient of restitution is expressed by the following equation [37]:

$$e = -\frac{v_p' - v_b'}{v_p - v_b} \tag{9}$$

where $v_p'$ and $v_b'$ are the velocities of the projectile and buffer after the collision. $v_p$ and $v_b$ are the velocities of the projectile and buffer before the collision. Substituting $v_p' = V_1$, $v_b' = v(0)$, $v_p = V_0$, and $v_b = 0$ into Eq. (9), we get

$$V_1 = v(0) - eV_0 \tag{10}$$

From Eqs. (8) and (10), $v(0)$ can be obtained as follows;

$$v(0) = \frac{(1+e)m_p}{m_p + m_b} V_0 \tag{11}$$

The theoretical velocity ($V_0 = \sqrt{2gh}$) was used as the initial velocity of the projectile in this study. In this study, the coefficient of restitution was determined from the motion histories of the projectile and the buffer during the experiment, as described later in section 3.

*2.3.2 Estimation of kinetic energy stored in water in a pipe from experimental data*

As a result of the kinetic energy of the buffer being transmitted to the water in the pipe, a certain amount of energy is stored in the water in the pipe. If the energy stored in the water in the pipe is $\Pi_f$, it can be represented by the sum of the kinetic energy of the water $K_f$, the work of the water $W_f$ owing to the pressure fluctuation, and the strain energy $U$ stored in the pipe:

$$\Pi_f = K_f + W_f + U \tag{12}$$

Energy is stored in the water in the pipe to the point reached by the pressure. Figure 3 presents a conceptual diagram of a mechanical model when a pressure wave propagates in the axial direction through the water in a pipe. The distance that the pressure wave has propagated at time $t$ is $c_\text{K}t$. If the cross-sectional area of the water is $A_\text{w}$ and the particle velocity in the vicinity of the interface with the buffer where the pressure wave is generated is $u(t)$ (Fig. 3(a)).

$$K_\text{f} = \int_0^t \frac{1}{2}\rho_\text{w} c_\text{K} A_\text{w} dt \cdot \{u(t)\}^2$$
$$= \frac{1}{2}\frac{A_\text{w}}{\rho_\text{w} c_\text{K}}\int_0^t \{p(t)\}^2 dt \tag{13}$$

$$W_\text{f} = \int_0^t p(t) A_\text{w} dt \cdot u(t)$$
$$= \frac{A_\text{w}}{\rho_\text{w} c_\text{K}}\int_0^t \{p(t)\}^2 dt \tag{14}$$

In the above, the relation $u(t) = p(t)/\rho_\text{w} c_\text{K}$ of the Joukowsky equation is used. Furthermore, $p(t)$ can be predicted from the hoop strain of the outer pipe wall using the following equation proposed by Tijsseling [38]:

$$p(t) = E_\text{tube}\frac{a}{R}\left(1 + \frac{1}{2}\frac{a}{R}\right)(\varepsilon_\phi - \varepsilon_{\phi,\text{out}}) - p_\text{out} \tag{15}$$

where $\varepsilon_{\varphi,\text{out}} = -(1-\nu)p_\text{out}/E_\text{pipe}$ is the hoop strain of the external pipe wall when $p = p_\text{out}$, and $p_\text{out}$ is the outer pressure. If the initial inner diameter of the pipe is $R_0$ and the average hoop strain of the outer pipe wall is $\overline{\varepsilon_\phi}$ (Fig. 3(b)),

$$U = \frac{\pi c_\text{K} a(a+2R_0)E_\text{tube}}{2}\overline{\varepsilon_\phi}^2 t \tag{16}$$

In this case, $\overline{\varepsilon_\phi}$ is the average strain in the part where the pressure wave propagates and can be expressed as follows using the circumferential strain $\varepsilon_{\phi,\text{int}}$ in the vicinity of the interface with the buffer where the pressure wave is generated:

$$\overline{\varepsilon_\phi} = \frac{\int_0^t \varepsilon_{\phi,\text{int}} dt}{c_K t} \quad (17)$$

Here, $U$ is in the order of $10^{-3}$ compared to $K_f$ and $W_f$, and it can be ignored because it is sufficiently small. From the above, the energy stored by the water can be obtained as follows:

$$\Pi_f \simeq \frac{3}{2} \frac{A_w}{\rho_w c_K} \int_0^t \{p(t)\}^2 dt \quad (18)$$

In this study, to obtain the energy expressed in Eq. (16), the pressure estimated with the hoop strain of the outer pipe wall measured at g4 was used, where g4 was located 10 mm from the buffer–water interface. The material properties used in the energy calculations in this study are listed in Table 2.

## 3 RESULTS AND DISCUSSION

### 3.1 Characteristics of wave propagation across solid–fluid interface

The images extracted from the movies recorded around the bottom of the buffer for each surface wettability condition are presented in Figs. 4–6. The origin of the time (0 ms) is the point where the strain was generated at the gauge directly beside the interface of the buffer and water (g3). With the untreated buffer surface condition, cavitation bubbles appeared from 0 ms in cycles of approximately 0.12 ms, although the water near the bottom of the buffer was continuously compressed by the buffer. The cavitation bubbles appeared as a result of the transmitted tensile stress wave propagating through the buffer into the water. The mechanism by which stress waves generated by the collision of the projectile and the buffer induce cavitation was described in a previous study [27]. It can be observed from Fig. 4 that the cavitation intensity became more severe with a larger drop height. This is due to the amplitude of the generated stress waves increased.

It should be noted that the response of the present experimental system depends on the length and the speed of sound of the projectile and buffer. They would change the period of repeated transmission of the stress waves into the water. As a result, the frequency of cavitation occurrence will change. Consequently, the energy transmitted to the pressure wave in the water will be affected. In the present experimental system, the natural frequencies of the projectile and the buffer which are owing to longitudinal stress wave propagation in the axial direction are 24.7 kHz and 8.6 kHz, respectively. During the experiment, cavitation appeared in the cycle of 1.2 ms, so the frequency of cavitation occurrence was 8.3 kHz, which is similar to the natural frequency of the buffer. Moreover, in the previous study, it was reported that the pressure wave speed in a bubbly-flow changes because the wave propagates while crushing the bubble [39]. In the present experiment, a similar phenomenon may occur close to the interface when stress waves are repeatedly transmitted into the water during cavitation occurrence.

In contrast to the condition of using an untreated buffer surface, few cavitation bubbles were observed with the plasma-treated Al foil tape (Fig. 6). The generation of cavitation bubbles was inhibited according to the decrease in the contact angle of the buffer edge. The cavitation generation was also inhibited with the untreated Al foil tape (Fig. 5). Thus, the attenuation of the transmitted wave amplitude owing to the tape had a strong effect on the cavitation generation, although the attenuation of the wave itself could be small.

In previous studies, the response of a solid–fluid interface on a water blast (wave propagation) was analyzed and modeled with cavitation occurrence [17,40]. Our experimental results clearly demonstrate that changing the surface wettability on the solid–fluid interface inhibited cavitation occurrence in the fluid. This provides new insight into the wave propagation across the solid-fluid interface; that is, the boundary condition between the solid and fluid at the solid–fluid interface may change the wave transmission behavior as a result of the change in the occurrence of cavitation.

Figures 7 and 8 show the representative data of the strain histories in the case with the projectile drop height of 200 mm measured at gauges g1–g11. Each line in Figs. 7 and 8 are displaced according to the distance from the interface of the buffer and water. The strain history measured at g1 corresponded to the time history of the incident waveform. Comparing Figs. 7(a), (b), and (c), the incident waves exhibited similar profiles; however, the amplitudes of the strain history measured at g2 were different. This corresponds to the effect of the wave reflected from the interface of the buffer and water with different buffer surface conditions. The strain histories in Fig. 8 indicate the pipe expansion owing to pressure wave propagation. The wave profile differed from the that of incident stress wave, possibly because of the difference in the wave speed between the buffer and water in the pipe, which caused the stress wave to be transmitted into the water continually, while being repeatedly reflected at the edges of the buffer, and the water to be compressed by the buffer motion [41].

The above experimental results demonstrated that the solid surface wettability affected the wave propagation in the FSI. New insight revealed that, on the solid–fluid interface, the boundary condition between the solid and fluid could change the wave transmission behavior as a result of the change in the cavitation intensity. Consequently, it can be said that a more accurate model for FSI or new methods to reduce the damage to structures by FSI can be proposed considering the solid surface condition at the interface.

**3.2 Evaluation of cavitation effect by energy transmission**

The effect of the cavitation intensity on the wave propagation with the difference in the solid surface wettability was evaluated using the energy transmission quantity. First, the buffer velocity was obtained using Eq. (6) to calculate the buffer kinetic energy. To obtain the velocity expressed by Eq. (6), the velocity change before and after the collision was obtained from the motion histories of the projectile and the buffer (Fig. 9), then the restitution coefficient was obtained by Eq. (9). The motion histories in Fig. 9 were extracted from high-speed video camera recordings using the same method as in previous studies [27]. The calculated coefficient was $e = 0.49$. To confirm the validity of the velocity obtained using Eq. (6), it was substituted into Eq. (4), converted into pressure, and compared with the pressure at the buffer–water interface measured in the previous experiment [27] (Fig. 10). A comparison was conducted with the condition of a projectile drop height of 100 mm. The pressure obtained by the theoretical analysis in this study agreed well with the averaged value of the measured data in the previous experiment. Therefore, the average transition tendency of the buffer motion could be predicted effectively.

The $K_{s,\text{lost}}$ values were determined using the obtained velocity of the buffer, and then compared with $\Pi_f$. Figure 11 presents the representative data of $K_{s,\text{lost}}$ and $\Pi_f$ over time for each condition with a drop height of 200 mm. The energy stored in the water in the pipe was small with poorer buffer surface wettability (a larger contact angle). The cavitation intensity from the buffer–water interface increased with poorer wettability conditions (Figs. 4–6). As a result, the energy transmitted to the water may have been consumed for cavitation formation.

When a stress wave is transmitted as a pressure wave from the buffer to the water in the pipe, an associated pressure wave with a profile similar to that of an underwater explosion is formed [42], owing to the difference in the speed of sound between the buffer and water. Because of the difference in the sound speed, a stress wave is repeatedly transmitted into the water by repeated reflections at the buffer edge, and the water is compressed by the buffer motion to form an associated pressure wave. However, this study revealed that energy is consumed by repeated cavitation inception when the associated wave is formed. According to Fig. 10, the duration of the associated wave was 4.2 ms in the present experiment. Hence, the energy stored in the water from 0 to 4.2 ms, which corresponded to the duration of the associated pressure wave, was compared with the lost kinetic energy of the buffer in the same period for each condition of the projectile drop height, as illustrated in Fig. 12. Comparing Fig. 12 with the flow features in Figs. 4 to 6, it can be observed that the more severe cavitation intensity with poorer wettability (a larger contact angle) could consume more energy for the pressure transmittance at every value of $v_0$. Figures 4 to 6 indicate that the cavitation intensity was more severe with larger tensile stress at a faster velocity (a higher drop height), as also evidenced in Fig. 12. As indicated in Fig. 12, for the conditions in which cavitation was observed, the difference between $K_{s,lost}$ and $\Pi_f$ appeared to increase as the initial velocity increased.

To clarify the relationship between the consumed energy and the wettability of the buffer–water interface with cavitation generation, the differences between $K_{s,lost}$ and $\Pi_f$ at the same initial velocity are compared in Fig. 13. It can be observed that the energy difference increased as the initial velocity increased with the larger contact angle (poorer wettability) conditions ($\theta = 93.2°$ and $77.1°$). In the condition of $\theta = 12.5°$, where cavitation was hardly observed, the energy difference scarcely changed, even if the initial velocity changed. Consequently, the effect of the cavitation inception on the wave propagation at the solid–fluid interface with FSI could be evaluated quantitatively by considering the energy transferred from the solid to the water in the pipe.

## 4 CONCLUSION

We attempted to evaluate the effect of the surface wettability of the solid medium on the wave propagation across the solid–fluid interface with FSI quantitatively in the case of longitudinal wave propagation vertically towards the interface. The surface condition of the solid medium was changed by attaching thin tapes with different wettabilities to the buffer surface. During the experiment, while the water was continuously compressed, cavitation bubbles appeared as a result of the transmitted tensile wave propagating across

This is an author-submitted article to the Journal of Pressure Vessel Technology, Transactions of the ASME. The published version is available in the ASME Digital Collection at https://doi.org/10.1115/1.4056438.the interface in a cycle. The generation of the cavitation bubbles was inhibited according to the decrease in the contact angle on the solid buffer surface; consequently, the characteristic of pressure transmission across the interface changed. Thus, it was demonstrated that the solid surface wettability affected the wave propagation in the FSI. New insight revealed that, on the solid–fluid interface, the boundary condition between the solid and fluid could change the wave transmission behavior as a result of the change in the cavitation intensity. Consequently, it can be said that a more accurate model for FSI or new methods to reduce the damage to structures by FSI can be proposed by taking into account the solid surface condition at the interface.

The effect of the cavitation intensity on the wave propagation with the difference in wettability of the solid surface was evaluated using the energy transmission quantity. The kinetic energy lost by the buffer was derived theoretically. Thereafter, the energy stored in the water in the pipe was estimated with the measured strain at the outer pipe wall and compared to the energy lost by the buffer. It appeared that more severe cavitation intensity with poorer wettability (a larger contact angle) consumed more energy of the pressure transmittance to the water. With the larger contact angle (poorer wettability) conditions, the difference between the energy lost by the buffer and that stored in the water increased as the initial buffer velocity increased. With improved wettability, where cavitation was hardly observed, the energy difference scarcely changed, even if the initial velocity changed. Consequently, the effect of the cavitation inception on the wave propagation across the solid–fluid interface with FSI could be quantitatively evaluated by considering the energy transferred from the solid to the water in the pipe.

**ACKNOWLEDGMENT**

We are grateful to Prof. Kikuo Kishimoto for the fruitful discussions.

**FUNDING**

This work was supported by the Japan Society for the Promotion of Science, Grants-in-Aid for Scientific Research [grant numbers JP26709001 and JP18K13662].

**NOMENCLATURE**

| | |
|---|---|
| $A_w$ | the cross-sectional area of the water, m² |
| $E_{pipe}$ | Young's modulus of the pipe material |
| $K$ | water volume modulus |
| $K_f$ | the kinetic energy of the water, J |
| $K_s(t)$ | the kinetic energy of the buffer at time $t$, J |
| $K_{s,lost}$ | kinetic energy lost by the buffer, J |
| $R$ | the representative radius of the pipe, m |
| $R_0$ | the initial inner diameter of the pipe, m |
| $U$ | strain energy stored in the pipe, J |
| $V_0$ | projectile velocity before the collision, m/s; $\left(\sqrt{2gh}\right)$ |
| $V_1$ | projectile velocity after the collision, m/s |
| $W_f$ | work of the water owing to the pressure fluctuation, J |
| $a$ | pipe wall thickness, m |
| $c_K$ | Korteweg velocity, m/s |
| $c_w$ | sound speed of the water, m/s |
| $g$ | gravity acceleration constant, m/s² |
| $h$ | drop height of the projectile, m |
| $m_b$ | buffer mass, kg |
| $m_p$ | projectile mass, kg |
| $p(t)$ | pressure in water at time $t$, Pa |
| $p_{out}$ | outer pressure, Pa |
| $t$ | time, s |
| $u(t)$ | particle velocity of water in the vicinity of the interface with the buffer, m/s |
| $v(t)$ | buffer velocity, m/s |
| $v_b$ | the velocity of the buffer before the collision, m/s; (= 0) |
| $v_p$ | the velocity of the projectile before the collision, m/s; (= $V_0$) |
| $v_b'$ | velocity of the buffer after the collision, m/s; (= $v(0)$) |
| $v_p'$ | the velocity of the projectile after the collision, m/s; (= $V_1$) |
| $\Pi_f$ | energy stored in the water in the pipe, J |
| $\varepsilon_{\phi,int}$ | circumferential strain in the vicinity of the interface with the buffer and water |
| $\varepsilon_{\varphi,out}$ | hoop strain of the external pipe wall when $p = p_{out}$; $\left(-(1-v)p_{out}/E_{pipe}\right)$ |
| $\overline{\varepsilon_\phi}$ | average hoop strain of the outer pipe wall in the part where the pressure wave propagates |
| $\rho_w$ | water density, kg/m³ |
| b | buffer |
| f | fluid |
| K | Korteweg |
| int | interface |
| lost | lost |
| out | out |
| p | projectile |
| pipe | pipe |
| s | solid |
| w | water |
| φ | hoop direction |
| 0 | initial |
| 1 | after the collision |
| FSI | Fluid–structure interaction |

**Figure Captions List**

Fig. 1    Schematic of experimental apparatus

Fig. 2    Water droplets on the solid surface for each condition: (a) untreated buffer surface, (b) untreated Al foil tape, and (c) plasma-treated Al foil tape

Fig. 3    Pressure wave propagating in the axial direction through water in a pipe

Fig. 4    Flow features at bottom of buffer with the untreated surface at a contact angle of 93.2° and with projectile drop heights of (a) 100 mm, (b) 200 mm, and (c) 300 mm

Fig. 5    Flow features at bottom of buffer with Al foil tape at a contact angle of 77.1° and with projectile drop heights of (a) 100 mm, (b) 200 mm, and (c) 300 mm

Fig. 6    Flow features at bottom of buffer with plasma-treated Al foil tape at a contact angle of 12.5° and with projectile drop heights of (a) 100 mm, (b) 200 mm, and (c) 300 mm

Fig. 7    Axial strain histories of buffer for each surface condition (representative data): (a) untreated buffer surface with a contact angle of 93.2°, (b) untreated Al foil tape with a contact angle of 77.1°, and (c) plasma-treated Al foil tape with a contact angle of 12.5°

Fig. 8    Hoop strain histories of pipe for each buffer surface condition (representative data): (a) untreated buffer surface with a contact angle of 93.2°, (b) untreated Al foil tape with a contact angle of 77.1°, and (c) plasma-treated Al foil tape with a contact angle of 12.5°

**Table Caption List**

Table 1          Contact angles and roughness of buffer surface

Table 2          Material properties for energy calculation

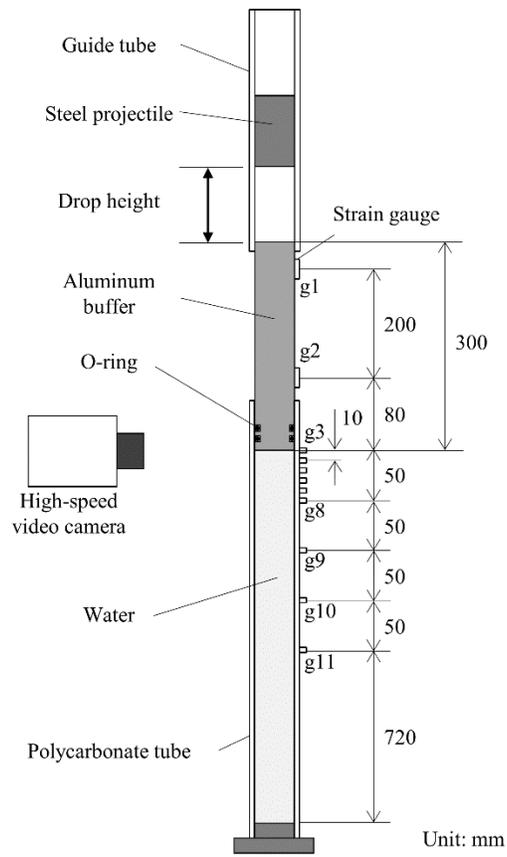

Fig. 1 Schematic of experimental apparatus

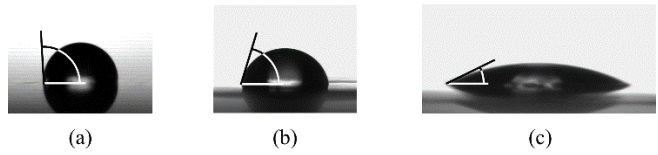

(a)  (b)  (c)

Fig. 2 Water droplets on the solid surface for each condition: (a) untreated buffer surface, (b) untreated Al foil tape, and (c) plasma-treated Al foil tape

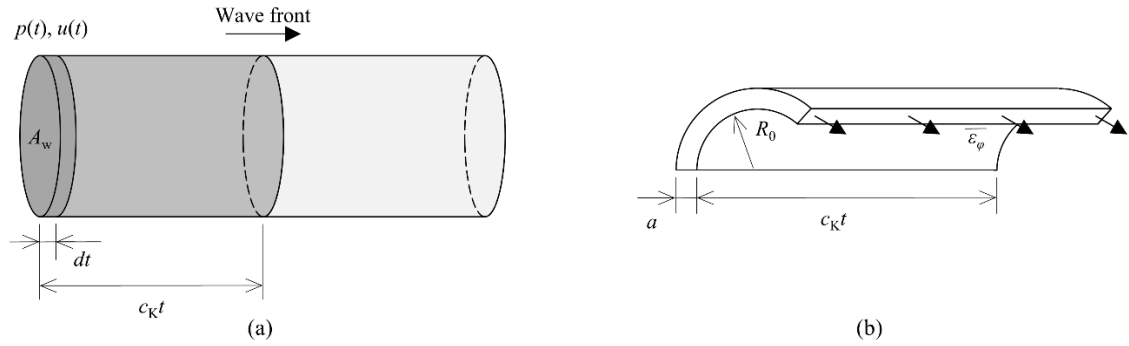

Fig. 3 Pressure wave propagating in the axial direction through water in a pipe

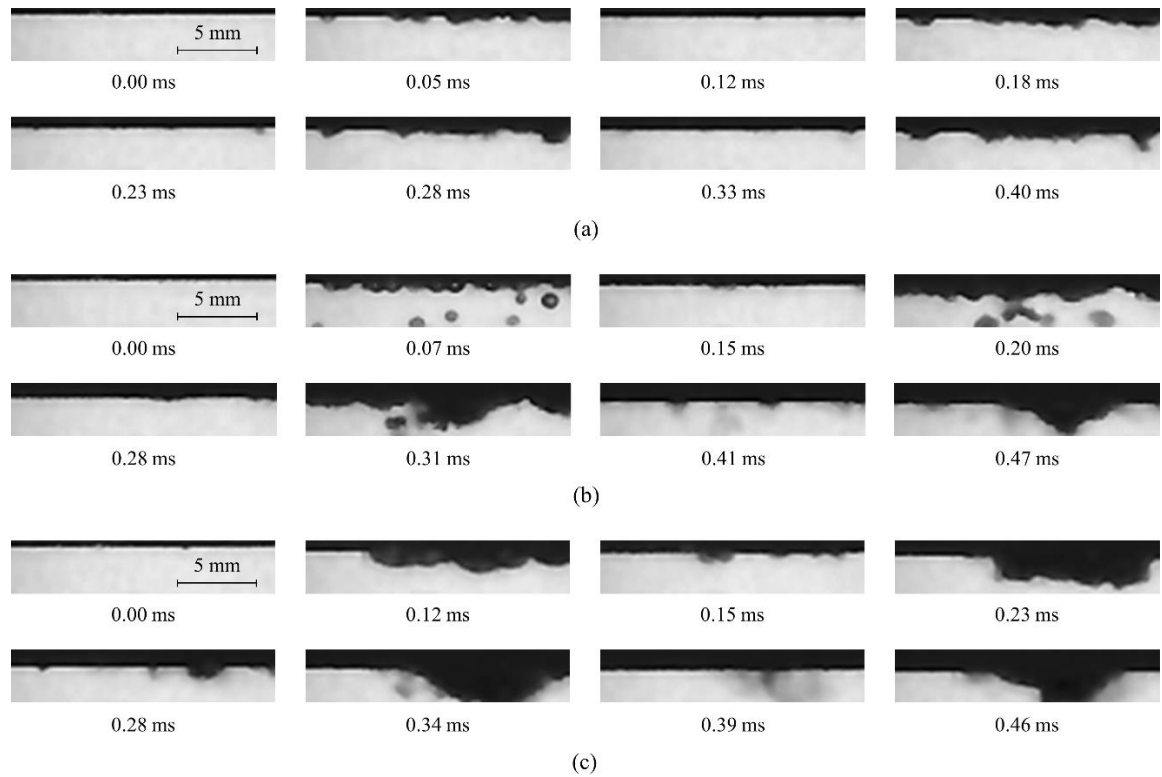

Fig. 4 Flow features at bottom of buffer with the untreated surface at a contact angle of 93.2° and with projectile drop heights of (a) 100 mm, (b) 200 mm, and (c) 300 mm

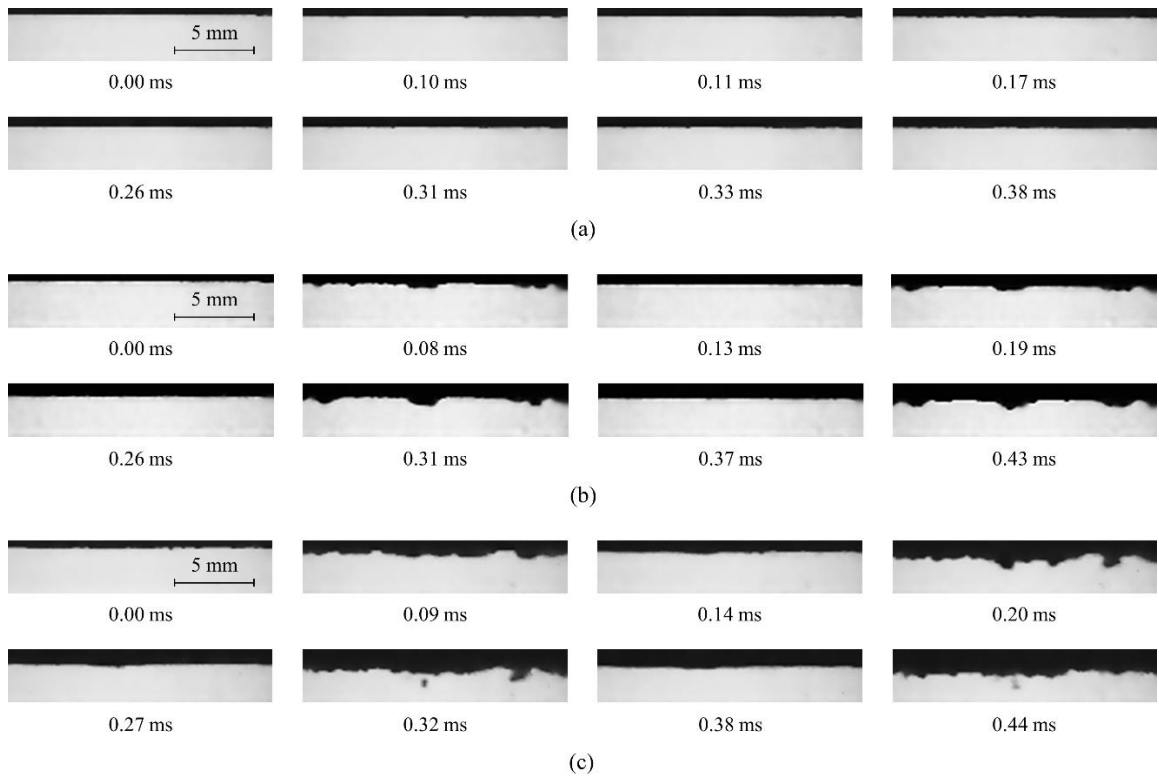

Fig. 5 Flow features at bottom of buffer with Al foil tape at a contact angle of 77.1° and with projectile drop heights of (a) 100 mm, (b) 200 mm, and (c) 300 mm

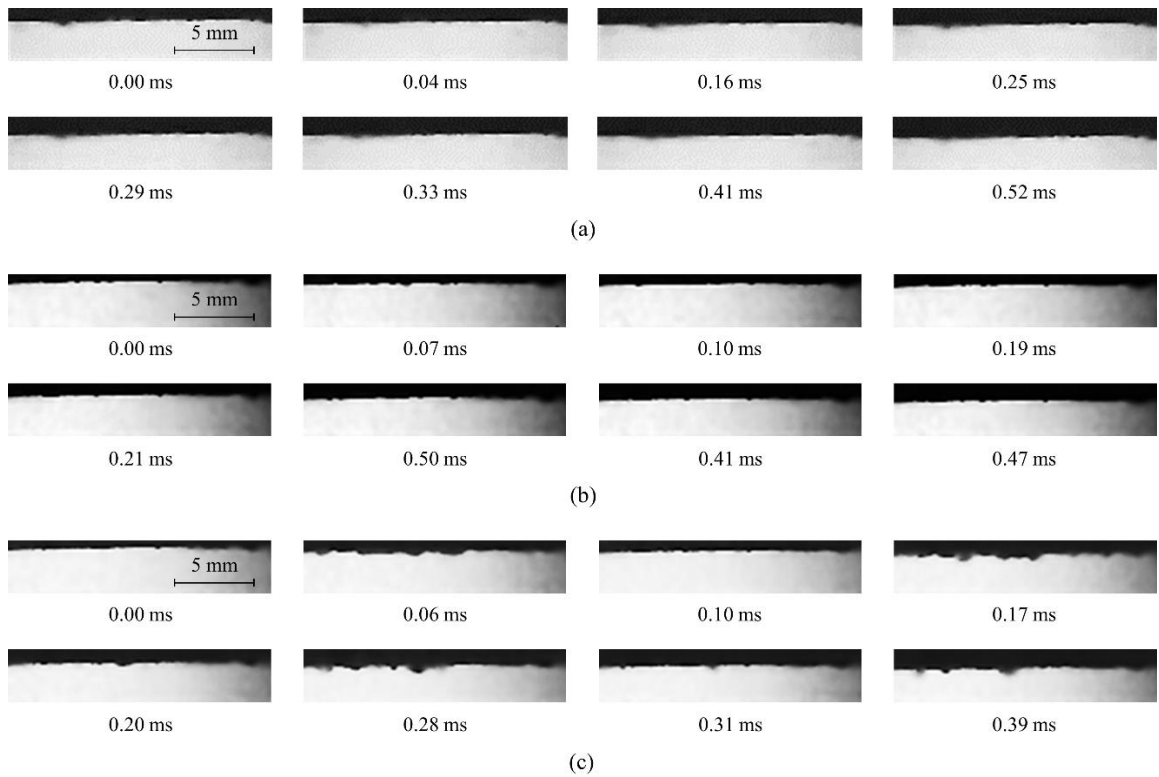

Fig. 6 Flow features at bottom of buffer with plasma-treated Al foil tape at a contact angle of 12.5° and with projectile drop heights of (a) 100 mm, (b) 200 mm, and (c) 300 mm

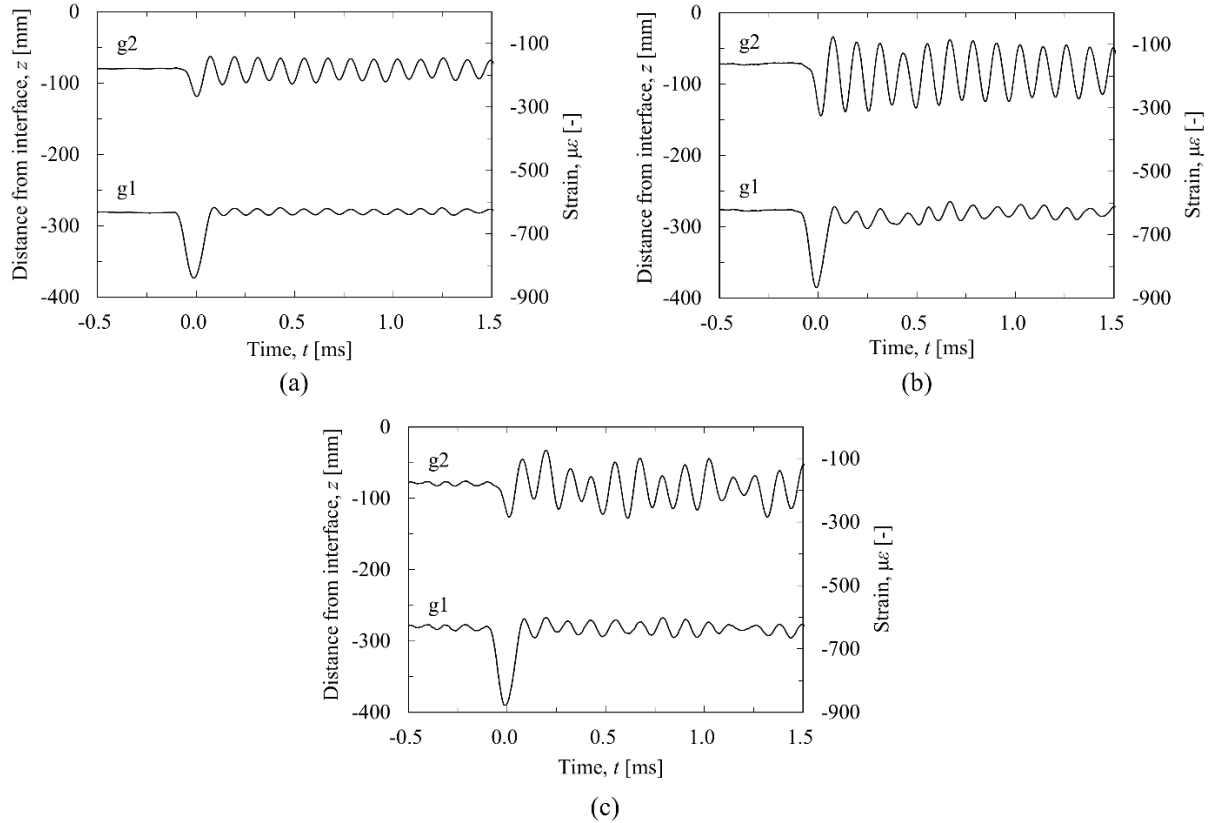

Fig. 7 Axial strain histories of buffer for each surface condition (representative data): (a) untreated buffer surface with a contact angle of 93.2°, (b) untreated Al foil tape with a contact angle of 77.1°, and (c) plasma-treated Al foil tape with a contact angle of 12.5°

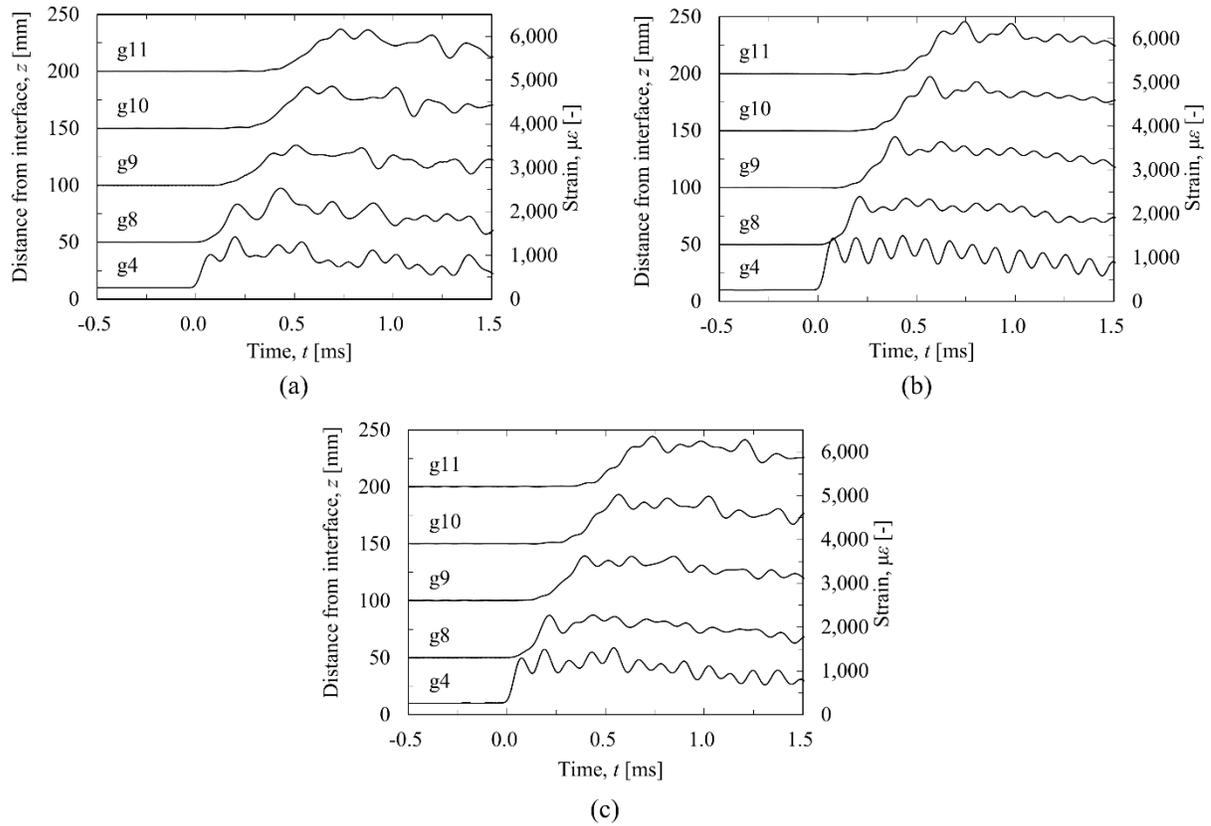

Fig. 8 Hoop strain histories of pipe for each buffer surface condition (representative data): (a) untreated buffer surface with a contact angle of 93.2°, (b) untreated Al foil tape with a contact angle of 77.1°, and (c) plasma-treated Al foil tape with a contact angle of 12.5°

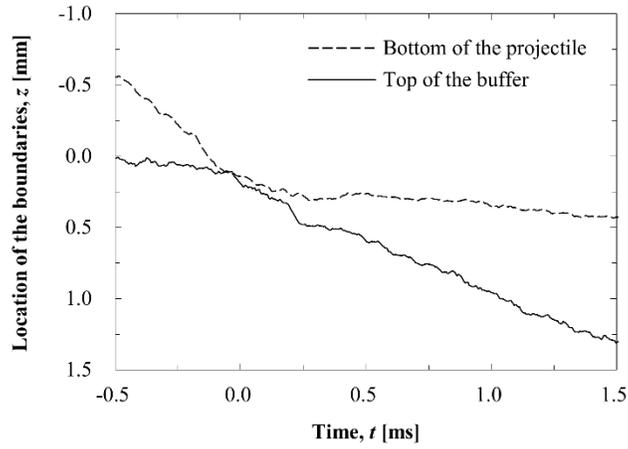

Fig. 9 Locations of buffer and projectile boundaries with untreated buffer surface and projectile drop height of 100 mm

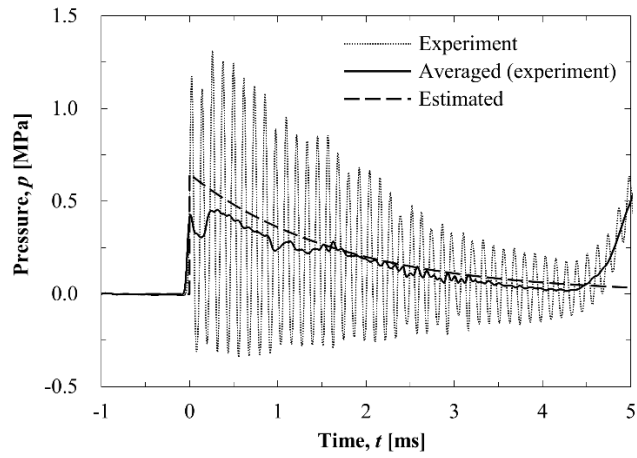

Fig. 10 Comparison of estimated pressure at buffer–water interface with measured data in the previous experiment

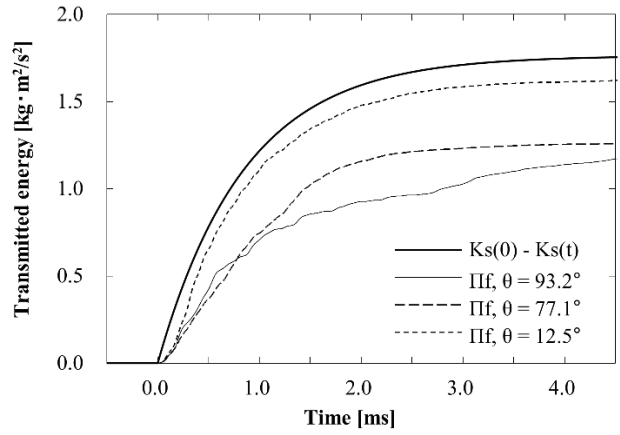

Fig. 11 Kinetic energy lost by buffer (theoretical value) and energy stored in water (experimental value)

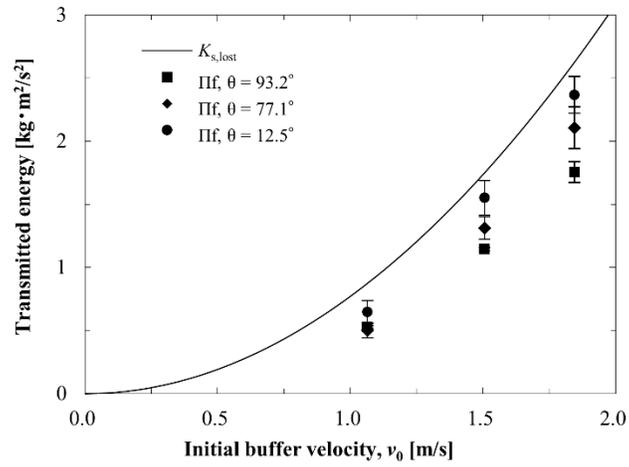

Fig. 12 Comparison of energy stored in water over entire synthetic wave formed in water in the pipe

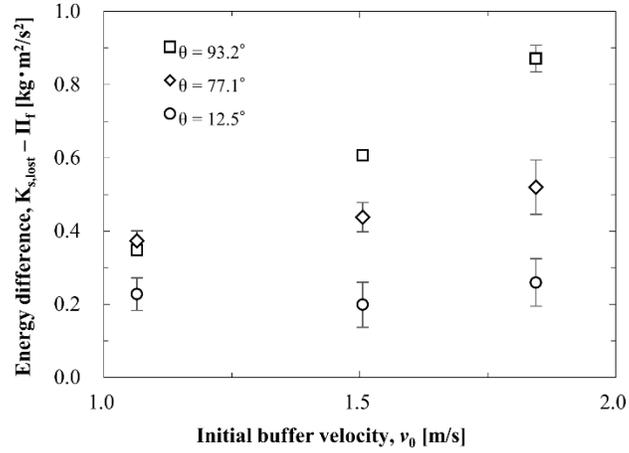

Fig. 13 Difference between energy lost by solid and energy stored in water for each condition

Table 1 Contact angles and roughness of buffer surface

| Buffer surface condition | Contact angle [deg] | The surface roughness Ra [μm] |
| --- | --- | --- |
| Untreated buffer surface | 93.2 (SE: 0.3) | 3.14 (SE: 0.16) |
| With untreated Al foil tape | 77.1 (SE: 1.9) | 2.46 (SE: 0.08) |
| With plasma-treated Al foil tape | 12.5 (SE: 1.8) | 2.89 (SE: 0.10) |

SE: standard error

Table 2 Material properties for energy calculation

| $m_b$ [kg] | $m_p$ [kg] | $\rho_w$ [kg/m³] | $c_w$ [m/s] | $K$ [GPa] | $E_{pipe}$ [GPa] | $p_{out}$ [MPa] |
|---|---|---|---|---|---|---|
| 1.56 | 1.62 | 1 000 | 1 483 | 2.06 | 2.45 | 0.101 |